\documentclass[a4paper,11pt]{article}
\usepackage{glossaries}
\usepackage[utf8]{inputenc}
\usepackage{pos}
\usepackage{graphicx}
\usepackage{amsthm, amsfonts, amsmath, mathrsfs}
\usepackage{bbm}
\usepackage{subcaption}
\usepackage{slashed}
\usepackage{physics}
\usepackage{xspace}
\usepackage{enumitem}

\newacronym{pionless}{EFT$_\slashed \pi$}{pionless effective field theory}
\newacronym{eft}{EFT}{effective field theory}
\newacronym{lecs}{LECs}{low energy coefficients}
\newacronym{qed}{QED}{quantum electrodynamics}
\newacronym{qcd}{QCD}{quantum chromodynamics}
\newacronym{latticeqcd}{LQCD}{lattice QCD}
\newacronym{nrqed}{NRQED}{nonrelativistic QED}
\newacronym{vnrqed}{vNRQED}{velocity NRQED}
\newacronym{nrqcd}{NRQCD}{nonrelativistic QCD}
\newacronym{ChiPT}{\ensuremath{\chi}PT}{chiral perturbation theory}
\newacronym{ChiEFT}{\ensuremath{\chi}EFT}{chiral effective field theory}
\newacronym{bsm}{BSM}{Beyond the Standard Model}
\newacronym{sm}{SM}{Standard Model}
\newacronym{NN}{\ensuremath{N\!N}\xspace}{nucleon-nucleon}
\newacronym{vRG}{vRG}{velocity renormalization group}
\newacronym{lo}{LO\ensuremath{_\slashed \pi}}{leading order}
\newacronym{nlo}{\ensuremath{\text{NLO}_{\slashed \pi}}}{next-to-leading order}
\newacronym{n2lo}{\ensuremath{\text{N}^2\text{LO}_{\slashed \pi}}}{next-to-next-to-leading order}
\newacronym{n3lo}{\ensuremath{\text{N}^3\text{LO}_{\slashed \pi}}}{next-to-next-to-next-to-leading order}
\newacronym{n4lo}{\ensuremath{\text{N}^4\text{LO}}}{next-to-next-to-next-to-next-to-leading order}
\newacronym{dimreg}{DimReg}{dimensional regularization}
\newacronym{av18}{AV18}{Argonne \ensuremath{v18}}
\newacronym{LLalpha}{\ensuremath{\text{LL}_\alpha}}{leading-logarithm-in \ensuremath{\alpha}}
\newacronym{NLLalpha}{\ensuremath{\text{NLL}_\alpha}}{next-to-leading-logarithm-in \ensuremath{\alpha}}
\newacronym{PDS}{PDS}{Power Divergence Subtraction}
\newacronym{BBN}{BBN}{big bang nucleosynthesis}

\newcommand{\oneS}{{{}^{1}\!S_0}}
\newcommand{\threeS}{{{}^{3}\!S_1}}
\newcommand{\NN}{\ensuremath{N\!N}\xspace}
\usepackage{ulem}

\title{Radiative corrections for the two-nucleon interaction in effective field theory
}

\author*[a]{Immo C. Reis}

\affiliation[a]{Institut f\"ur Kernphysik and PRISMA$^+$ Cluster of Excellence, Johannes Gutenberg-Universit\"at,\\ 55128 Mainz, Germany}

\emailAdd{imreis@students.uni-mainz.de}

\abstract{Radiative corrections are playing an increasingly significant role in low-energy nuclear physics. We investigate the influence of photons as explicit degrees of freedom within nuclear EFT on the deuteron binding energy, charge form factor, and the radiative capture process \( np \to d\gamma \) using the velocity renormalization group. In each case, evolving the subtraction velocity to a characteristic nucleon velocity in the bound state induces percent-level modifications in the relevant observables. These findings indicate that electromagnetic corrections represent a non-negligible source of uncertainty in current few-body calculations.
}

\FullConference{The 11th International Workshop on Chiral Dynamics (CD2024)\\
 26-30 August 2024\\
Ruhr University Bochum, Germany\\}

\begin{document}
\maketitle

\section{Introduction}  
Nuclear interactions are formulated within \gls{eft}~\cite{weinbergNuclearForcesChiral1990,weinbergEffectiveChiralLagrangians1991,weinbergThreeBodyInteractions1992,machleidtChiralEffectiveField2011a,epelbaumModernTheoryNuclear2009a,
hammerNuclearEffectiveField2020a,
kaplanNucleonNucleonScattering1996, kaplanNewExpansionNucleonnucleon1998, kaplanTwoNucleonSystems1998, vankolckEffectiveFieldTheory1999b}, wherein nucleons—and potentially pions and $\Delta$ isobars—constitute the relevant degrees of freedom. The \gls{lecs} of the nuclear Hamiltonian and currents are constrained by experimental data. 
In conventional nuclear \gls{eft}, photons are not treated as explicit degrees of freedom; instead, \gls{qed} effects are absorbed into the numerical values of the \gls{lecs}. 
However, for modern precision tests of the Standard Model, it is necesseary to obtain a thorough theoretical understanding of radiative corrections from \gls{qed} and their impact on the \gls{lecs} of nuclear \gls{eft}s.

Simultaneously, baryon–baryon calculations in \gls{latticeqcd} have undergone significant advancements in recent years~\cite{Illa_2021,Gongyo_2020}, raising the prospect of determining \gls{lecs} directly from \gls{qcd}. When combined with a nuclear \gls{eft} that incorporates photons as explicit degrees of freedom, this development will enable parameter-free \gls{sm} predictions for nuclear observables.
Accordingly, a systematic delineation of \gls{qed} and \gls{qcd} effects through the inclusion of photons as explicit dynamical fields in nuclear \gls{eft} is both timely and necessary.

In this work, we investigate the impact of radiative corrections on the neutron–proton system, in particular on the deuteron binding energy, the charge form factor, and the radiative capture process $np\to d\gamma$. The latter two observables, probed through muonic atom spectroscopy~\cite{pohlSizeProton2010, pohlLaserSpectroscopyMuonic2016,antogniniMuonicAtomSpectroscopyImpact2022} and measurements of light element abundances from \gls{BBN}~\cite{wagonerSynthesisElementsVery1967, steigmanPrimordialNucleosynthesisPrecision2007, ioccoPrimordialNucleosynthesisPrecision2009, cyburtBigBangNucleosynthesis2016, aghanimPlanck2018Results2020, cookeOnePercentDetermination2018, burlesSharpeningPredictionsBigBang1999}, serve as stringent tests of the \gls{sm} and necessitate a rigorous treatment of \gls{qed} radiative corrections.

\section{Framework}
In order to treat both \gls{qcd} and \gls{qed} in the neutron-proton system, we use a combination of \gls{pionless} and \gls{nrqed} along with the \gls{vRG}.
In \gls{pionless} it is typical to count powers of the momentum $p$, but in \gls{nrqed} powers of velocity $v = p/M_N \ll m_\pi/M_N$,  where $m_\pi$ is the pion mass and $M_N$ is the nucleon mass, are counted. 
The (energy, momentum) scales of the \gls{eft} are hard $(m_\pi,m_\pi)$, soft $(M_N v, M_N v)$, ultrasoft $(M_N v^2, M_N v^2)$, and potential $(M_N v^2, M_N v)$.
The nonrelativistic four-momentum of the nucleon is decomposed as \cite{lukeRenormalizationGroupScaling2000}
    \begin{equation}
            \label{eq:background:nrqed:nucleon_momentum}
        P = (0, \vb p) + (k_0, \vb k) \, ,
    \end{equation}
where $\vb p \sim M_N v$ is the soft component of the momentum and $k \sim M_N v^2$ is the residual four-momentum on the ultrasoft scale.
The degrees of freedom in the theory consist of potential nucleons, soft photons, and ultrasoft photons.
The nucleon fields are written as $N_{\vb p} (x)$ where $\vb p$ is a soft label, $x$ is the Fourier conjugate of the residual momentum $k$, and $N$ is an isodoublet of the proton and neutron.
The photon field is split into a soft mode, $A^\mu_p (k)$ with soft label four-momentum $p$ and a residual four-momentum $k$, and an ultrasoft mode, $A^\mu$.
The potential photons can be integrated out because they are far off-shell; their effects are encoded in the coefficients of four-nucleon operators.
Additionally, soft nucleons could be included in the theory, but they can just as well be integrated out for the same reason.

The \gls{vRG} was formulated in Refs.~\cite{lukeRenormalizationGroupScaling2000,manoharRenormalizationGroupCorrelated2000} by introducing two separate but correlated scales in dimensional regularization.
There is the ultrasoft scale $\mu_U$ and the soft scale $\mu_S$, and they are related according to $\mu_U = \mu_S^2/M_N$.
A subtraction velocity $\nu$ can be introduced according to $\mu_S = M_N \nu$ so that both $\mu_U$ and $\mu_S$ are simultaneously described by a single $\nu$.
The soft and ultrasoft anomalous dimensions of a generic coupling $C$ are obtained through
\cite{lukeRenormalizationGroupScaling2000,manoharRenormalizationGroupCorrelated2000}
    \begin{align}
        \mu_U \frac{d C}{d \mu_U} & = \gamma_U \, , \\
        \mu_S \frac{d C}{d \mu_S} & = \gamma_S \, ,
    \end{align}
which lead to the \gls{vRG} equation
    \begin{align}
        \nu \frac{d C}{d \nu} & = \gamma_S + 2 \gamma_U \, .
    \end{align}
In \gls{nrqed}, the electron mass is integrated out, so the fine-structure constant $\alpha$ does not run.
In the \NN system, the electron is kept as a light degree of freedom, so $\alpha$ still runs in \gls{pionless}.
Here, the anomalous dimensions have a dual expansion,
    \begin{align}
        \gamma & = \sum_{m=-1, n=0} \gamma^{(m, n)} \, ,
    \end{align}
where $\gamma^{(m, n)} \sim v^m \alpha^n$.

In \gls{pionless}, the nucleons interact with one another through potentials.
These potentials will be renormalized by ultrasoft interactions.
The neutron-proton potential is expanded as
    \begin{align}
        V_{pn} & = \sum_{v=-1} \sum_{\vb p', \vb p} V^{(v)}_{abcd} (\vb p', \vb p) p^\dagger_{\vb p', a} p_{\vb p, b} n^\dagger_{- \vb p', c} n_{-\vb p, d} \, ,
    \end{align}
where $v$ tracks the order in the velocity expansion of each coefficient, and $a$, $b$, $c$, and $d$ are spin indices for the proton and neutron.
The \gls{lo}, \gls{nlo}, and \gls{n2lo} potential coefficients in the S-wave are given by\footnote{Our definition of $C_4$ is a linear combination of $C_4 + \tilde C_4$ that appears in most of the \gls{pionless} literature, see Ref.~\cite{richardsonRadiativeCorrectionsRenormalization2024}.}
    \begin{align}
        V^{(-1)}_{abcd} & = C_{0}^{(\threeS)} P^{(1)}_{ab, cd} + C_{0}^{(\oneS)} P^{(0)}_{ab, cd} \, , \\
        V^{(0)}_{abcd} & = \frac{1}{2} \left( \vb p'^2 + \vb p^2 \right) \left[ C_{2}^{(\threeS)} P^{(1)}_{ab, cd} + C_{2}^{(\oneS)} P^{(0)}_{ab, cd} \right] \, , \\
        V^{(1)}_{abcd} & = \frac{1}{4} \left( \vb p'^2 + \vb p^2 \right)^2 \left[ C_{4}^{(\threeS)} P^{(1)}_{ab,cd} + C_{4}^{(\oneS)} P^{(0)}_{ab, cd} \right] \, ,
    \end{align}
where the projection operators are given by
    \begin{align}
        P^{(1)}_{ab, cd} & = \frac{1}{4} \left( 3 \delta_{ab} \delta_{cd} + \sigma^i_{ab} \sigma^i_{cd} \right) \, , \\
        P^{(0)}_{ab, cd} & = \frac{1}{4} \left( \delta_{ab} \delta_{cd} - \sigma^i_{ab} \sigma^i_{cd} \right) \, ,
    \end{align}
and the superscripts indicate the total spin projection.

\subsection{Renormalization of the strong potential}

\begin{figure}
    \centering
    
    \includegraphics[width=0.9\textwidth]{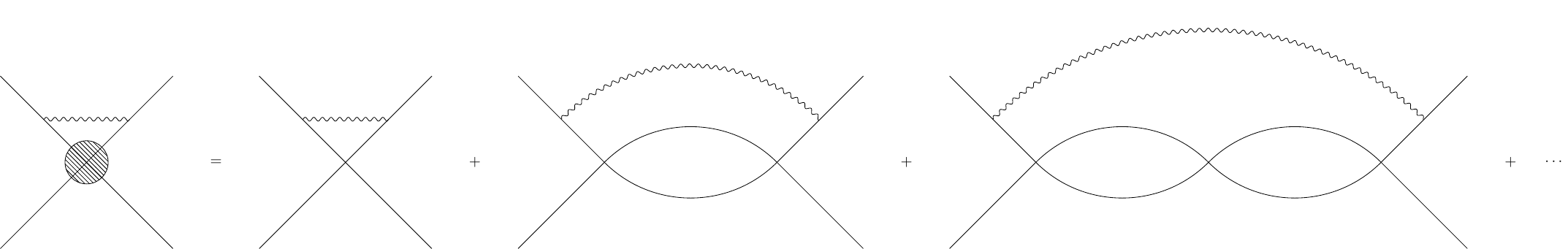}%

\caption{$O(\alpha/v)$ diagrams that contribute to the anomalous dimension of the potential.}
\label{fig:diagrams}
\end{figure}

In Refs.~\cite{richardsonRadiativeCorrectionsRenormalization2024,secondpaper}, we calculated the leading anomalous dimensions of $C_0$, $C_2$ and $C_4$ in the $\threeS$ and $\oneS$ channels. In Feynman gauge, the dominant contribution, which is $O(\alpha/v)$, comes from an $A_0$ photon coupled to the proton on both the incoming and outgoing lines with insertions of the $C_0$ potential. In the bubble series, a graph with $l = 2j$ bubbles, where $j$ is a strictly positive integer, requires a counterterm that renormalizes the $2j$-derivative potential.
For example, the diagram with 2 \NN bubbles, the third diagram on the right in Fig.~\ref{fig:diagrams}, renormalizes the $V^{(0)}$ potential. The $V^{(-1)}$ potential is not renormalized at this order.

The solutions of the corresponding \gls{vRG} equations are
    \begin{align}
        C_2(\nu) & = C_2\left( \frac{m_\pi}{M_N} \right) - \frac{27}{8} \left( \frac{M_N}{4 \pi} \right)^2 C_0^3 \log \left( \frac{\alpha(M_N \nu^2)}{\alpha(m_\pi^2/M_N)} \right) \, , \\
        C_4(\nu) & = C_4\left( \frac{m_\pi}{M_N} \right) + \frac{15}{4} \left( \frac{M_N}{4 \pi} \right)^4 C_0^5 \log \left( \frac{\alpha(M_N \nu^2)}{\alpha(m_\pi^2/M_N)} \right) \, ,
    \end{align}
    where the running \gls{qed} coupling in minimal subtraction is 
    \begin{align}
        \alpha(\mu) & = \frac{\alpha_{\text{OS}}}{1 - \frac{2 \alpha_{\text{OS}}}{3 \pi} \log \left( \frac{\mu}{m_e} \right)} \, ,
    \end{align}
where $\alpha_{\text{OS}} = 1/137$ and $m_e = 511$ keV, with OS indicating that this is the fine structure constant in the on-shell scheme.
The values of the \gls{lecs} at the matching scale $\nu = m_\pi/M_N$ where the pions are in principle integrated out are fixed according to,
    \begin{align}
        C^{(s)}_{0} & = \frac{4 \pi a_s}{M_N} \, , \label{eq:C0_running} \\
        C^{(s)}_{2} \left( \frac{m_\pi}{M_N} \right) & = \frac{2 \pi a_s^2 r_s}{M_N} \, , \label{eq:C2_running} \\
        C^{(s)}_{4} \left( \frac{m_\pi}{M_N} \right) & = \frac{4 \pi}{M_N} a_s^3 \left( \frac{1}{4} r_s^2 + \frac{P_s}{a_s} \right) \, , \label{eq:C4_running}
    \end{align}
where $a_s$, $r_s$ and $P_s$ are the scattering length, effective range and shape parameter in channel $s$, respectively. As a proxy for a pure \gls{qcd} result we take their numerical values from Ref.~\cite{wiringaAccurateNucleonnucleonPotential1995} with $\alpha = 0$ and Ref.~\cite{deswartLowEnergyNeutronProtonScattering1995a}. Ideally, they would be taken from \gls{latticeqcd} instead.

Working at \gls{n2lo}+ \gls{LLalpha}, all \gls{qed} contributions up to that order and the leading logarithmic contributions of higher orders are included via renormalization of the strong potential. Explicit photon diagrams start to contribute at \gls{n3lo}~\cite{richardsonRadiativeCorrectionsRenormalization2024}. It should be noted that varying the subtraction velocity $\nu$ in results provides an idea of the effect of higher order logarithms that would arise in perturbation theory, i.e., it can be used to assess higher order \gls{qed} contributions\footnote{See, e.g., Ref.~\cite{Cohen:2019wxr} for a pedagogical discussion.}.

\section{Results}
After having obtained the leading anomalous dimensions of the potential, we can now turn our attention to the renormalization group analysis of deuteron properties.
\subsection{Deuteron binding energy}
\begin{figure}[t]
    \centering
    \includegraphics[width=0.8\columnwidth]{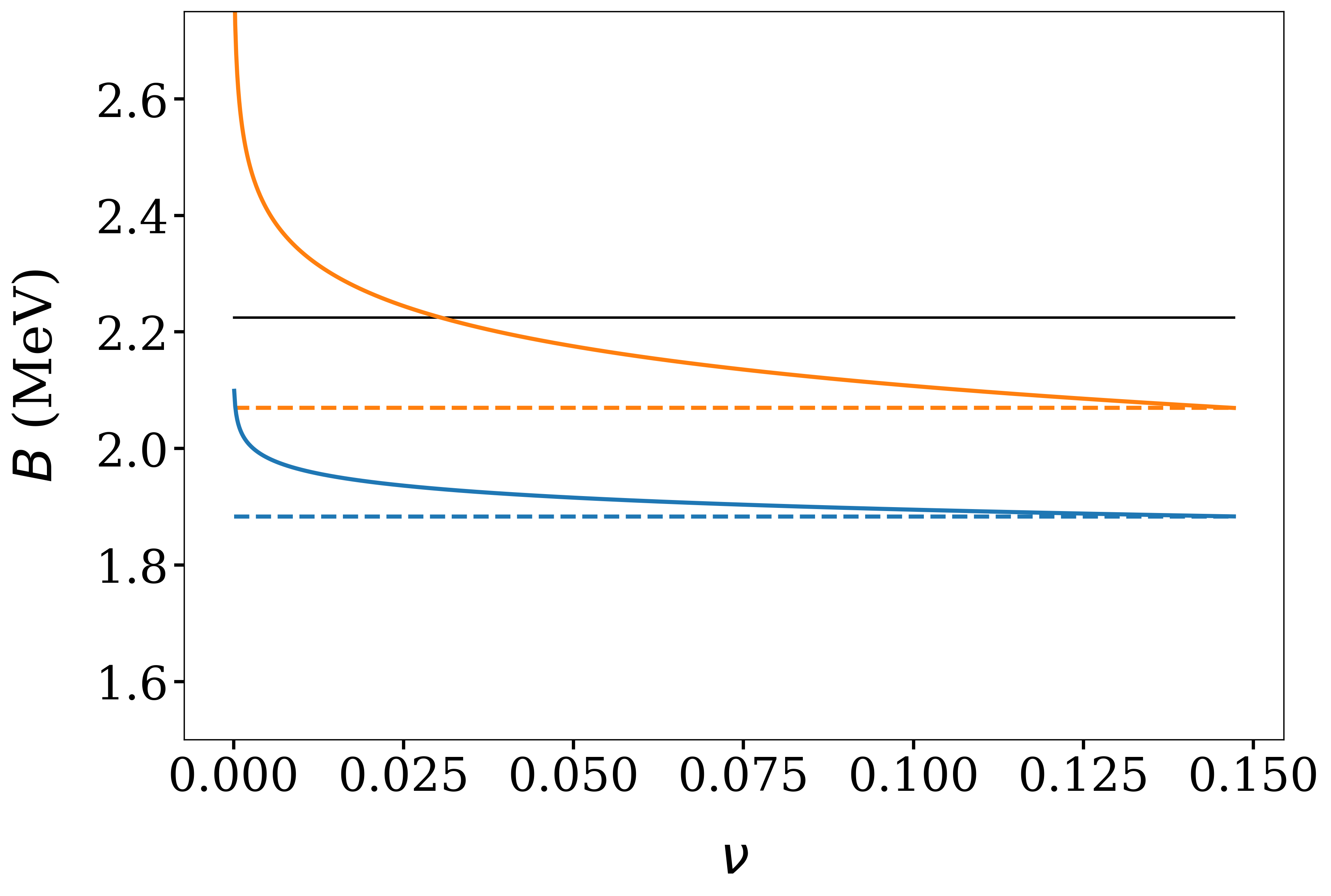}
    \caption{The deuteron binding energy as a function of the subtraction velocity. The solid black line is the experimental value. The blue (orange) dashed line is the fixed order \gls{nlo} (\gls{n2lo}) result while the blue (orange) solid line is the renormalization group improved \gls{nlo} (\gls{n2lo}) result. Figure from Ref.~\cite{richardsonRadiativeCorrectionsRenormalization2024}.}
    \label{fig:deuteron_binding}
\end{figure}

We show the deuteron binding energy as a function of the subtraction velocity at \gls{nlo}+\gls{LLalpha} and \gls{n2lo}+\gls{LLalpha} in Fig.~\ref{fig:deuteron_binding}.
First, we can compare the values of the renormalization group improved binding energies at each order in the \gls{eft} to the values at the hard scale $\nu = m_\pi/M_N$.
When the subtraction velocity is $\nu \approx 0.04$ (corresponding to momenta around $38$ MeV), there is a shift in the binding energy of about $2.5\%$ at \gls{nlo}.
At \gls{n2lo}, the improvement shifts the binding energy by about $7\%$.
Moreover, the improvement at \gls{n2lo} causes the predicted binding energy to intersect the experimental value $B = 2.224575$ MeV around $\nu \approx 0.04$.

\subsection{Deuteron form factors}

\begin{figure}
    \centering    \includegraphics[width=0.75\textwidth]{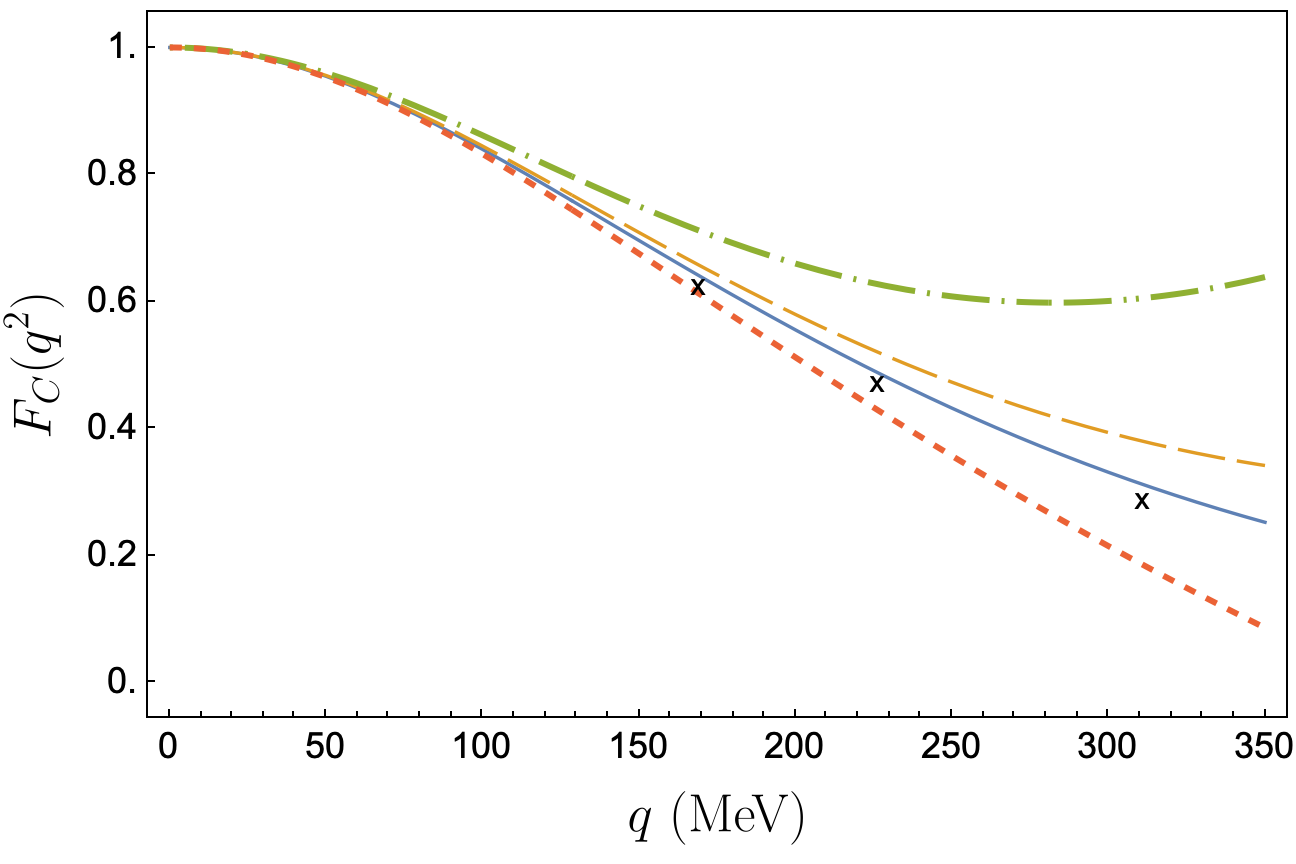}
    \caption[The solid (blue), dashed (orange), and dot-dashed (green) lines show the RG improved \gls{n2lo} charge form factor at $\nu = 0.1, \, 0.08,$ and $0.04$, respectively. The dotted (pink) line shows the \gls{n2lo} calculation in the $Z$-parameterization. The black crosses are data.]{The solid (blue), dashed (orange), and dot-dashed (green) lines show the RG improved \gls{n2lo} charge form factor at $\nu = 0.1, \, 0.08,$ and $0.04$, respectively. The dotted (pink) line shows the \gls{n2lo} calculation in the $Z$-parameterization \cite{ lenskyForwardDoublyvirtualCompton2021}. The black crosses are data taken from Ref.~\cite{abbottPhenomenologyDeuteronElectromagnetic2000}. Figure from Ref.~\cite{secondpaper}.}
    \label{fig:charge_FF}
\end{figure}

The result for the charge form factor is shown in Fig.~\ref{fig:charge_FF}.
For comparison, the pink dashed line shows the result of the $Z$-parameterization calculation at \gls{n2lo} \cite{ lenskyForwardDoublyvirtualCompton2021} and the experimental data is taken from Ref.~\cite{abbottPhenomenologyDeuteronElectromagnetic2000}.
We vary the subtraction velocity from $\nu = 0.1$ to $0.04$ as this provides an idea of the effect of higher order logarithms that would arise in perturbation theory.
At $\nu = 0.1$, the RG improved prediction is in rough agreement with the experimental data for fairly large momentum transfers.
As $\nu$ is lowered even further, the theoretical prediction moves further away from the data, which might be improved in the next logarithmic order.
However, the disagreement between the theoretical and experimental results is strongest for $q > m_\pi$ where \gls{pionless} breaks down.
\begin{figure}
    \centering
    \includegraphics[width=0.75\textwidth]{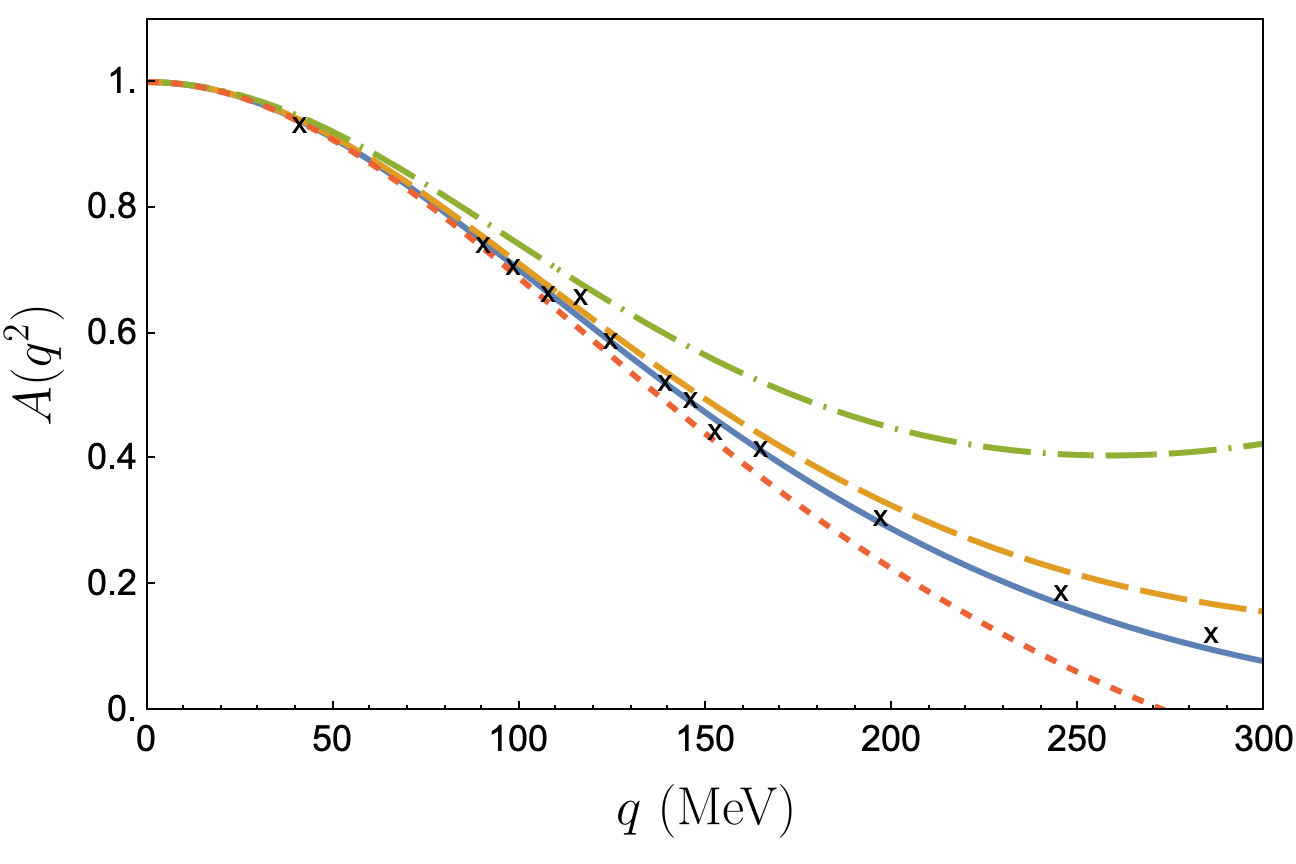}
    \caption[The solid (blue), dashed (orange), and dot-dashed (green) lines show the RG improved \gls{n2lo} charge form factor at $\nu = 0.1, \, 0.08,$ and $0.04$, respectively. The dotted (pink) line shows the \gls{n2lo} calculation in the $Z$-parameterization. The black crosses are data.]{The solid (blue), dashed (orange), and dot-dashed (green) lines show the RG improved \gls{n2lo} charge form factor at $\nu = 0.1, \, 0.08,$ and $0.04$, respectively. The dotted (pink) line shows the \gls{n2lo} calculation in the $Z$-parameterization \cite{ lenskyForwardDoublyvirtualCompton2021}. The black crosses are data taken from Ref.~\cite{simonElasticElectricMagnetic1981}. Figure from Ref.~\cite{secondpaper}.}
    \label{fig:charge_A}
\end{figure}

In Fig.~\eqref{fig:charge_A}, we display the \gls{vRG} improved quantity,
    \begin{align}
        A(q^2) & = F_C^2(q^2) + \frac{2}{3} \eta F_M^2(q^2) + \frac{8}{9} \eta^2 F_Q^2(q^2) \, ,
    \end{align}
where $\eta = -q^2/4 M_d^2$, $F_M$ is the magnetic form factor, and $F_Q$ is the quadrupole form factor.
At the order we are working, we only need the magnetic form factor at \gls{lo} \cite{kaplanPerturbativeCalculationElectromagnetic1999a},
    \begin{align}
        \frac{e}{2 M_d} F^{(0)}_M(q^2) & = \kappa_0 \frac{e}{M_N} F_C^{(0)}(q^2) \, .
    \end{align}
We compare the results to the analogous calculations in the $Z$-parameterization \cite{lenskyForwardDoublyvirtualCompton2021} at \gls{n2lo} and the data from Ref.~\cite{simonElasticElectricMagnetic1981}.
The features of the plot are identical to those in Fig.~\ref{fig:charge_FF}.
Again, we see excellent agreement with the data for $\nu = 0.1$ and $0.08$ for a wide range of momentum transfers.

From the form factor, we can also examine the conventional definition of the deuteron charge radius, 
    \begin{align}
        \langle r_d^2 \rangle_C & = - 6 \frac{d F_C(q^2)}{d q^2} \big|_{q^2 = 0} \, .
    \end{align}
At $\nu = m_\pi/M_N$, corresponding to no \gls{qed} in the theory, the theoretical charge radius $r_{\text{th}}(m_\pi/M_N) = \sqrt{\langle r_d^2 \rangle_C} = 2.15(6)$ fm, where we have assigned a $3\%$ error to account for the truncation of the \gls{eft}.
After the \gls{vRG} improvement, the charge radius at $\nu = 0.1$ is $r_{\text{th}}(0.1) = 2.11(6)$ fm.
Again, we have assigned a naive 3\% error assuming that the \gls{vRG} improvement does not strongly modify the truncation error of the \gls{eft}.
Despite this uncertainty, the \gls{vRG} has shifted the central value by around 2\%. Our result agrees within the \gls{pionless} uncertainty for $\nu \in [0.06, 0.1]$ with the precise experimental result of the CREMA collaboration \cite{pohlLaserSpectroscopyMuonic2016}, $r_{\text{CREMA}} = 2.12562(13)_{\text{exp}} (77)_{\text{th}}$.
Therefore, the summation of \gls{qed} logarithms through the \gls{vRG} seems to play an important role in reproducing the charge radius without recourse to the Z-parameterization, which subsumes corrections to all orders in $\alpha$ in the \gls{lecs}.

\subsection{Radiative neutron capture}

Lastly, we consider the radiative capture process $np \to d \gamma$, which is of fundamental importance to \gls{BBN}. This reaction is the first step in the BBN network as it is the reaction producing primordial deuterium. A rigorous quantification of this reaction's theory errors is therefore imperative.
Simulations of \gls{BBN}~\cite{pisantiPArthENoPEPublicAlgorithm2008, pitrouPrecisionBigBang2018,burlesSharpeningPredictionsBigBang1999} use theory predictions from Refs.~\cite{rupakPrecisionCalculationNp2000,andoRadiativeNeutronCapture2006} as input since the available experimental data are scarce and do not come with competitive errors. 
Reference~\cite{rupakPrecisionCalculationNp2000} quotes a 1\% uncertainty in the energy regime relevant for BBN. 
In the following, we will investigate the sensitivity of this process to \gls{qed} driven radiative corrections.

\begin{figure}
    \centering
    \begin{subfigure}[]{0.49\textwidth}
    		\label{fig:suzuki}
        \centering
        \includegraphics[width=\textwidth]{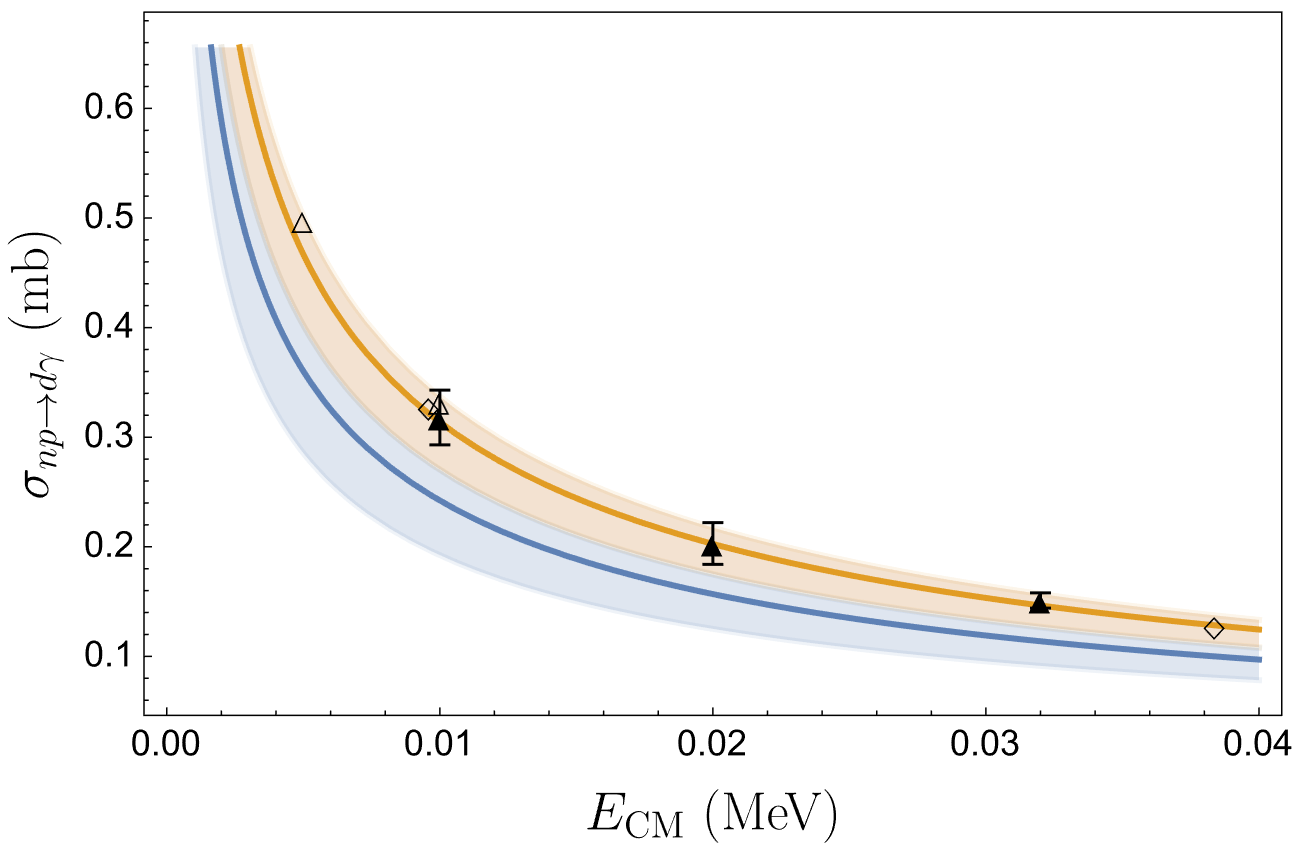}
        \caption{}
    \end{subfigure}
    \begin{subfigure}[]{0.49\textwidth}
    		\label{fig:nagai}
        \centering
        \includegraphics[width=\textwidth]{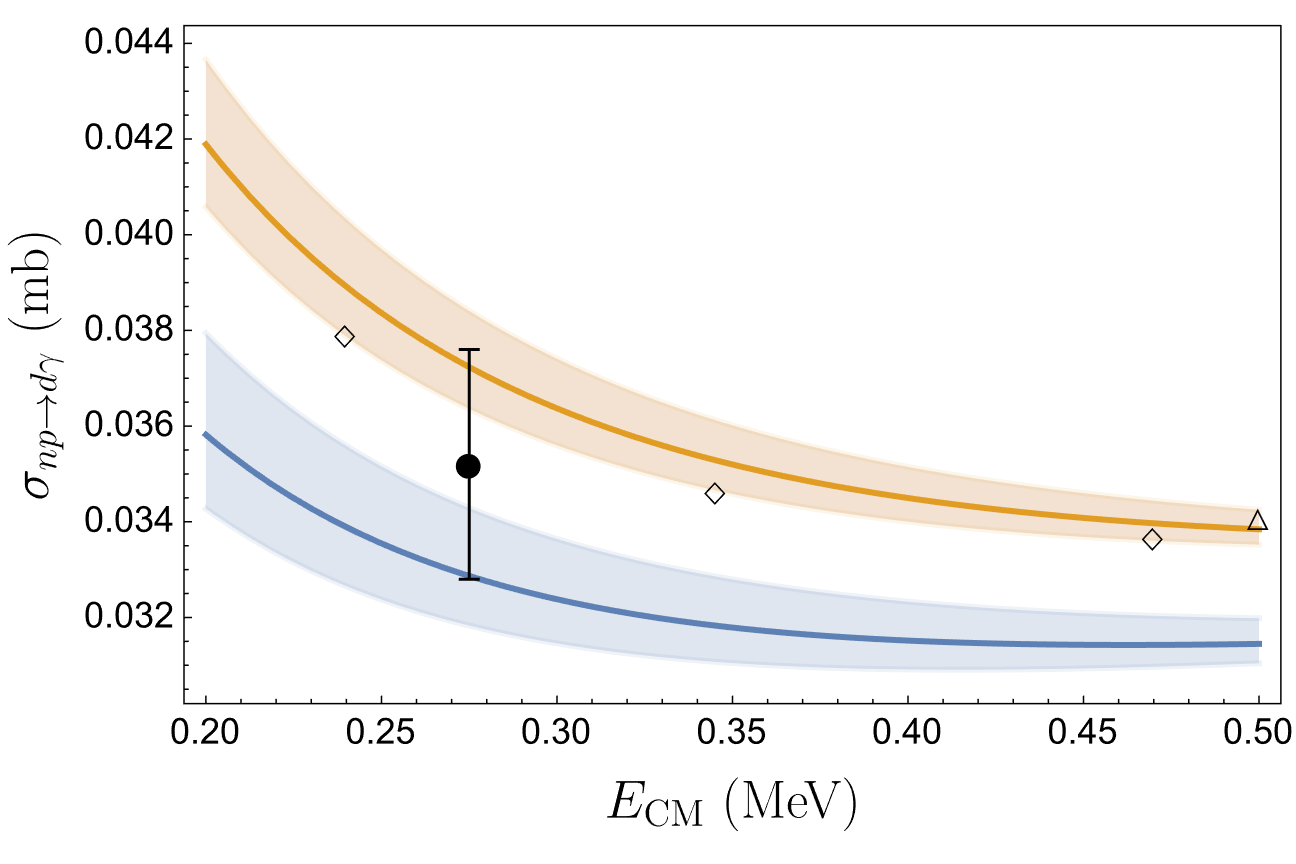}
        \caption{}
    \end{subfigure}
    \caption[Renormalization group improved cross section without $\tilde L_1$ in blue and with $\tilde L_1$ in orange.]{Renormalization group improved cross section without $\tilde L_1$ in blue and with $\tilde L_1$ in orange. The open triangles are the \gls{pionless} result of Ref.~\cite{rupakPrecisionCalculationNp2000}, and the open diamonds are the \gls{ChiEFT} result of Ref.~\cite{acharyaGaussianProcessError2022}.The experimental data come from Ref. \cite{suzukiFirstMeasurementPn1995} (upward triangles) and \cite{nagaiMeasurementReactionCross1997} (circle). Figure from Ref.~\cite{secondpaper}.} \label{fig:npdgamma_bins}
\end{figure}

In Fig.~\ref{fig:npdgamma_bins} we compare our results at \gls{nlo}+\gls{LLalpha} with and without the LEC $\tilde L_1$\footnote{We also show the results without $\tilde L_1$ as they provide a parameter free prediction. $\tilde L_1$ is usually fitted to the thermal neutron capture cross section data from Ref.~\cite{coxProtonthermalNeutronCapture1965} and therefore includes \gls{qed} effects to all orders. The results with $\tilde L_1$ are therefore, strictly speaking, not truly \gls{nlo}+\gls{LLalpha}. See Ref.~\cite{secondpaper} for details.} to the \gls{pionless} \cite{rupakPrecisionCalculationNp2000}, \gls{ChiEFT} results \cite{acharyaGaussianProcessError2022} and experimental data~\cite{ suzukiFirstMeasurementPn1995, nagaiMeasurementReactionCross1997}.
In particular, the central values of the data from Ref.~\cite{suzukiFirstMeasurementPn1995} are almost reproduced with the presence of $\tilde L_1$ and $\nu = 0.05$.
In this region, running from $\nu = m_\pi/M_N$ to 0.05 induces a shift in the cross section of 6-8\%.
Both curves are within the error of the single data point from Ref.~\cite{nagaiMeasurementReactionCross1997}, and the \gls{vRG} evolution leads to a change in the cross section of about 3\%.
At higher energies, the variation from the \gls{vRG} is negligible, so the higher order corrections in the \gls{pionless} power counting should be more important.

We find reasonable agreement with the \gls{pionless} \cite{rupakPrecisionCalculationNp2000} and \gls{ChiEFT} results \cite{acharyaGaussianProcessError2022}.
However, we note that Ref.~\cite{rupakPrecisionCalculationNp2000} states that electromagnetic effects, in particular the potential-like interaction coupling the proton and neutron magnetic moments, are included implicitly by making use of the experimental scattering length in the $\oneS$ channel.
A similar statement could be made regarding the \gls{ChiEFT} calculation since the parameters of the potential are generally fit to \NN scattering data.
In our approach, these effects are included via the renormalization group and go beyond the use of potential-like electromagnetic interactions.
Both Ref.~\cite{rupakPrecisionCalculationNp2000} and Refs.~\cite{acharyaGaussianProcessError2022, acharyaUncertaintyQuantificationElectromagnetic2023} find sub-percent level precision, which is important for comparison with experiment and astrophysical observations and for validation of the general \gls{eft} approach.
However, detecting new physics in this process will require a clean separation of different strong and electroweak effects. 

The renormalization group analysis presented here suggests that $np \to d \gamma$ is sensitive to \gls{qed} corrections to the \NN interaction at the few-percent level, which seems to be an unquantified source of uncertainty in existing calculations.

\section{Conclusion and outlook}
Here, we have presented an analysis of the role of radiative corrections in the neutron-proton interaction. We use \gls{eft} techniques to organize and delineate the role of different strong and electromagnetic effects in a systematic expansion.

We presented a \gls{vRG} analysis of the deuteron binding energy, deuteron charge form factor, charge radius and the radiative capture process $np \to d \gamma$. Working at \gls{n2lo}+\gls{LLalpha}, the \gls{vRG} shifts the deuteron binding energy at the few-percent level towards the experimental value and improves the description of form factor data as well as the deuteron charge radius prediction. 

For $np \to d \gamma$ at \gls{nlo}+\gls{LLalpha}, our analysis is in general agreement with available experimental data as well as other calculations in \gls{pionless} \cite{chenBigbangNucleosynthesis1999,rupakPrecisionCalculationNp2000} and \gls{ChiEFT} \cite{acharyaGaussianProcessError2022}.
However, to our knowledge, this is the first approach to incorporate the effects of radiative corrections through the renormalization group.
Electromagnetic effects are still implicitly included by fitting the LEC $\tilde L_1$ to the measured thermal neutron capture cross section \cite{coxProtonthermalNeutronCapture1965}.

The scale variation in our \gls{vRG} analysis indicates however, that this process is sensitive to higher order \gls{qed} corrections at the level of a few percent.
Thus, \gls{qed} corrections appear to constitute an unquantified source of uncertainty in the accepted theoretical prediction of the cross section. Extending our analysis to \gls{NLLalpha} would help increase the precision of the results.
Furthermore, a reassessment of thermal neutron capture from lattice \gls{qcd} would help disentangle the strong and electromagnetic effects of this process even more.

Our findings suggest that similar few-percent-level corrections may affect other few-body systems and reactions, such as proton-proton fusion, potentially exceeding the estimates in a recent study \cite{combesRadiativeCorrectionsProtonproton2024}.
\section*{Acknowledgements}
The author would like to thank the Mainz Physics Academy for financial support and Thomas Richardson for feedback on the manuscript. This work was supported in part by the Deutsche Forschungsgemeinschaft (DFG) through the Cluster of Excellence ``Precision Physics, Fundamental Interactions, and Structure of Matter'' (PRISMA${}^+$ EXC 2118/1) funded by the DFG within the German Excellence Strategy (Project ID 39083149).


\begin{thebibliography}{99}
\bibitem{weinbergNuclearForcesChiral1990}Weinberg, S. Nuclear Forces from Chiral Lagrangians. {\em Phys.Lett.}. \textbf{B251} pp. 288-292 (1990)
\bibitem{weinbergEffectiveChiralLagrangians1991}Weinberg, S. Effective Chiral Lagrangians for Nucleon - Pion Interactions and Nuclear Forces. {\em Nucl.Phys.}. \textbf{B363} pp. 3-18 (1991)
\bibitem{weinbergThreeBodyInteractions1992}Weinberg, S. Three Body Interactions among Nucleons and Pions. {\em Phys.Lett.}. \textbf{B295} pp. 114-121 (1992)
\bibitem{machleidtChiralEffectiveField2011a}Machleidt, R. \& Entem, D. Chiral Effective Field Theory and Nuclear Forces. {\em Physics Reports}. \textbf{503}, 1-75 (2011,6)
\bibitem{epelbaumModernTheoryNuclear2009a}Epelbaum, E., Hammer, H. \& Meißner, U. Modern Theory of Nuclear Forces. {\em Reviews Of Modern Physics}. \textbf{81}, 1773-1825 (2009,12)
\bibitem{hammerNuclearEffectiveField2020a}Hammer, H., König, S. \& Van Kolck, U. Nuclear Effective Field Theory: Status and Perspectives. {\em Reviews Of Modern Physics}. \textbf{92}, 025004 (2020,6)
\bibitem{kaplanNucleonNucleonScattering1996}Kaplan, D., Savage, M. \& Wise, M. Nucleon - Nucleon Scattering from Effective Field Theory. {\em Nucl.Phys.}. \textbf{B478} pp. 629-659 (1996)
\bibitem{kaplanNewExpansionNucleonnucleon1998}Kaplan, D., Savage, M. \& Wise, M. A New Expansion for Nucleon-Nucleon Interactions. {\em Phys.Lett.}. \textbf{B424} pp. 390-396 (1998)
\bibitem{kaplanTwoNucleonSystems1998}Kaplan, D., Savage, M. \& Wise, M. Two Nucleon Systems from Effective Field Theory. {\em Nucl.Phys.}. \textbf{B534} pp. 329-355 (1998)
\bibitem{vankolckEffectiveFieldTheory1999b}Van Kolck, U. Effective Field Theory for Short-Range Forces. {\em Nuclear Physics A}. \textbf{645}, 273-302 (1999,1)

\bibitem{Illa_2021} Illa, M., Beane, S., Chang, E., Davoudi, Z., Detmold, W., Murphy, D., Orginos, K., Parreño, A., Savage, M., Shanahan, P., Wagman, M. \& Winter, F. Low-energy scattering and effective interactions of two baryons at $m_\pi \sim 450$ MeV from lattice quantum chromodynamics. {\em Physical Review D}. \textbf{103} (2021).

\bibitem{Gongyo_2020}Gongyo, S., Sasaki, K., Miyamoto, T., Aoki, S., Doi, T., Hatsuda, T., Ikeda, Y., Inoue, T. \& Ishii, N. d⁎(2380) dibaryon from lattice QCD. {\em Physics Letters B}. \textbf{811} pp. 135935 (2020,12)


\bibitem{pohlSizeProton2010}Pohl, R. et al. The Size of the Proton. {\em Nature}. \textbf{466}, 213-216 (2010,7)
\bibitem{antogniniMuonicAtomSpectroscopyImpact2022}Antognini, A. et al. Muonic-Atom Spectroscopy and Impact on Nuclear Structure and Precision QED Theory. (arXiv,2022,10)
\bibitem{pohlLaserSpectroscopyMuonic2016}Pohl, R. et al. Laser Spectroscopy of Muonic Deuterium. {\em Science}. \textbf{353}, 669-673 (2016,8)

\bibitem{wagonerSynthesisElementsVery1967}Wagoner, R., Fowler, W. \& Hoyle, F. On the Synthesis of Elements at Very High Temperatures. {\em The Astrophysical Journal}. \textbf{148} pp. 3 (1967,4)
\bibitem{steigmanPrimordialNucleosynthesisPrecision2007}Steigman, G. Primordial Nucleosynthesis in the Precision Cosmology Era. {\em Annual Review Of Nuclear And Particle Science}. \textbf{57}, 463-491 (2007,11)
\bibitem{ioccoPrimordialNucleosynthesisPrecision2009}Iocco, F., Mangano, G., Miele, G., Pisanti, O. \& Serpico, P. Primordial Nucleosynthesis: From Precision Cosmology to Fundamental Physics. {\em Physics Reports}. \textbf{472}, 1-76 (2009,3)
\bibitem{cyburtBigBangNucleosynthesis2016}Cyburt, R., Fields, B., Olive, K. \& Yeh, T. Big Bang Nucleosynthesis: Present Status. {\em Reviews Of Modern Physics}. \textbf{88}, 015004 (2016,2)
\bibitem{aghanimPlanck2018Results2020} Aghanim, N. et al. Planck 2018 Results - VI. Cosmological Parameters. {\em Astronomy \& Astrophysics}. \textbf{641}, A6 (2020).
\bibitem{burlesSharpeningPredictionsBigBang1999}Burles, S., Nollett, K., Truran, J. \& Turner, M. Sharpening the Predictions of Big-Bang Nucleosynthesis. {\em Physical Review Letters}. \textbf{82}, 4176-4179 (1999,5)
\bibitem{cookeOnePercentDetermination2018}Cooke, R., Pettini, M. \& Steidel, C. One Percent Determination of the Primordial Deuterium Abundance*. {\em The Astrophysical Journal}. \textbf{855}, 102 (2018,3)





\bibitem{lukeRenormalizationGroupScaling2000}Luke, M., Manohar, A. \& Rothstein, I. Renormalization Group Scaling in Nonrelativistic QCD. {\em Physical Review D}. \textbf{61}, 074025 (2000,3)
\bibitem{manoharRenormalizationGroupCorrelated2000}Manohar, A., Soto, J. \& Stewart, I. The Renormalization Group for Correlated Scales: One-Stage versus Two-Stage Running. {\em Physics Letters B}. \textbf{486}, 400-405 (2000,8)


\bibitem{richardsonRadiativeCorrectionsRenormalization2024}Richardson, T. \& Reis, I. Radiative Corrections and the Renormalization Group for the Two-Nucleon Interaction in Effective Field Theory. {\em Few-Body Systems}. \textbf{65}, 79 (2024,8)
\bibitem{secondpaper}Richardson, T. \& Reis, I. Renormalization group analysis of electromagnetic properties of the deuteron.  (2025), https://arxiv.org/abs/2502.17634




\bibitem{wiringaAccurateNucleonnucleonPotential1995}Wiringa, R., Stoks, V. \& Schiavilla, R. Accurate Nucleon-Nucleon Potential with Charge-Independence Breaking. {\em Physical Review C}. \textbf{51}, 38-51 (1995,1)
\bibitem{deswartLowEnergyNeutronProtonScattering1995a}De Swart, J., Terheggen, C. \& Stoks, V. The Low-Energy Neutron-Proton Scattering Parameters and the Deuteron. (arXiv,1995,9)
\bibitem{Cohen:2019wxr}Cohen, T. As Scales Become Separated: Lectures on Effective Field Theory. {\em PoS}. \textbf{TASI2018} pp. 011 (2019)
\bibitem{lenskyForwardDoublyvirtualCompton2021}Lensky, V., Blin, A. \& Pascalutsa, V. Forward Doubly-Virtual Compton Scattering off an Unpolarised Deuteron in Pionless Effective Field Theory. {\em Physical Review C}. \textbf{104}, 054003 (2021,11)

\bibitem{abbottPhenomenologyDeuteronElectromagnetic2000}Abbott, D. et al. Phenomenology of the Deuteron Electromagnetic Form Factors. {\em The European Physical Journal A - Hadrons And Nuclei}. \textbf{7}, 421-427 (2000,3)


\bibitem{kaplanPerturbativeCalculationElectromagnetic1999a}Kaplan, D., Savage, M. \& Wise, M. A Perturbative Calculation of the Electromagnetic Form Factors of the Deuteron. {\em Physical Review C}. \textbf{59}, 617-629 (1999,2)

\bibitem{simonElasticElectricMagnetic1981}Simon, G., Schmitt, \. \& Walther, V. Elastic Electric and Magnetic E-d Scattering at Low Momentum Transfer. {\em Nuclear Physics A}. \textbf{364}, 285-296 (1981,7)



\bibitem{pisantiPArthENoPEPublicAlgorithm2008}Pisanti, O., Cirillo, A., Esposito, S., Iocco, F., Mangano, G., Miele, G. \& Serpico, P. PArthENoPE: Public Algorithm Evaluating the Nucleosynthesis of Primordial Elements. {\em Computer Physics Communications}. \textbf{178}, 956-971 (2008,6)
\bibitem{pitrouPrecisionBigBang2018}Pitrou, C., Coc, A., Uzan, J. \& Vangioni, E. Precision Big Bang Nucleosynthesis with Improved Helium-4 Predictions. {\em Physics Reports}. \textbf{754} pp. 1-66 (2018,9)



\bibitem{rupakPrecisionCalculationNp2000} Rupak, G. Precision Calculation of $Np \rightarrow d\gamma$ Cross Section for Big-Bang Nucleosynthesis. {\em Nuclear Physics A}. \textbf{678}, 405-423 (2000).
\bibitem{andoRadiativeNeutronCapture2006}Ando, S., Cyburt, R., Hong, S. \& Hyun, C. Radiative Neutron Capture on a Proton at Big-Bang Nucleosynthesis Energies. {\em Physical Review C}. \textbf{74}, 025809 (2006,8)
\bibitem{chenBigbangNucleosynthesis1999} Chen, J. \& Savage, M. $n p\rightarrow d\gamma$ for Big-Bang Nucleosynthesis. {\em Physical Review C}. \textbf{60}, 065205 (1999).


\bibitem{acharyaGaussianProcessError2022} Acharya, B. \& Bacca, S. Gaussian Process Error Modeling for Chiral Effective-Field-Theory Calculations of $Np \leftrightarrow d\gamma$ at Low Energies. {\em Physics Letters B}. \textbf{827}, 137011 (2022).



\bibitem{coxProtonthermalNeutronCapture1965}Cox, A., Wynchank, S. \& Collie, C. The Proton-Thermal Neutron Capture Cross Section. {\em Nuclear Physics}. \textbf{74}, 497-507 (1965,12)

\bibitem{suzukiFirstMeasurementPn1995}Suzuki, T. \& Others First Measurement of a p(n, Gamma)d Reaction Cross Section between 10 and 80 keV. {\em The Astrophysical Journal}. \textbf{439} pp. L59 (1995,2)
\bibitem{nagaiMeasurementReactionCross1997} Nagai, Y., Suzuki, T., Kikuchi, T., Shima, T., Kii, T., Sato, H. \& Igashira, M. Measurement of $^1\text{H}(n, \gamma) ^2\text{H}$ Reaction Cross Section at a Comparable $M1/E1$ Strength. {\em Physical Review C}. \textbf{56}, 3173-3179 (1997).




\bibitem{acharyaUncertaintyQuantificationElectromagnetic2023}Acharya, B., Bacca, S., Bonaiti, F., Li Muli, S. \& Sobczyk, J. Uncertainty Quantification in Electromagnetic Observables of Nuclei. {\em Frontiers In Physics}. \textbf{10} (2023,1)


\bibitem{combesRadiativeCorrectionsProtonproton2024}Combes, E., Mereghetti, E. \& Platter, L. Radiative Corrections to Proton-Proton Fusion in Pionless Effective Field Theory. {\em Physical Review C}. \textbf{110}, L041001 (2024,10)
\end{thebibliography}
\end{document}